\documentclass[prb,twocolumn,a4paper,showkeys]{revtex4-1} 

\usepackage{bm}
\usepackage{amsmath}

\def\a{\alpha}
\def\b{\beta}
\def\d{\delta}
\def\D{\Delta}
\def\e{\epsilon}

\def\l{\lambda}

\def\ff{\psi}
\def\g{\gamma}

\def\o{\omega}

\def\s{\sigma}
\def\ra{\rightarrow}

\begin{document}

\title{Penetration of fast projectiles into resistant media:\\ from 
macroscopic to subatomic projectiles}

\author{Jos\'e Gaite}
\affiliation{
Applied Physics Dept.,
ETSIAE,
Universidad Polit\'ecnica de Madrid,\\ E-28040 Madrid, Spain}

\email[Electronic address: ]{jose.gaite@upm.es}
\date{July 21, 2017}

\begin{abstract}
The penetration of a fast projectile into a resistant medium is 
a complex process that is suitable for simple modeling, in which 
basic physical principles can be profitably employed.
This study connects two different domains: the fast motion of
macroscopic bodies in resistant media and the interaction 
of charged subatomic particles with matter at high energies, which
furnish the two limit cases of the problem of penetrating projectiles of 
different sizes. These limit cases actually have overlapping applications; 
for example, in space physics and technology.
The intermediate or mesoscopic domain finds application in atom cluster 
implantation technology. Here it is shown that the penetration 
of fast nano-projectiles is ruled by a slightly 
modified Newton's inertial quadratic force, namely, $F \sim v^{2-\b}$,
where $\b$ vanishes as the inverse of projectile diameter.
Factors essential to penetration depth are ratio of projectile to medium density and projectile shape.
\end{abstract}


\maketitle

\section{Introduction}
\label{intro}

The analytical study of the resistance to the motion of projectiles 
begins with 
{\em Book Two} of Newton's {\em Principia}, entitled {\em The motion of
  bodies (in resisting mediums).}\cite{Newt}
Other classics have studied this subject,
which has obvious applications, for example, military applications.
In this regard, we can mention the classic treaty on gunnery
by Robins.\cite{Robins} His work was continued by Euler.\cite{Euler} Later,
Poncelet\cite{Poncelet} and Resal\cite{Resal} deduced formulas for
penetration depth that are still in use.
The simplest formula for penetration depth is due to Gamow\cite{Gamow} but he attributed it to Newton.\cite{Wiki,Saslow} 
Of course, the various aspects of the
impact and penetration of projectiles in resistant
media are treated in several modern books and reviews.\cite{Bronshten,Zukas,Kinslow,Iskander,Ruiz-S,Ben,Chinamen}

Newton's theory of resistance focuses on fluids,\cite{Newt} but 
it can be applied to the motion of {\em fast} projectiles in 
any media.\cite{Gamow} 
Projectiles with velocities about 1 km/s are fairly normal, while
larger velocities, about 10 km/s, 
are typical in space and, in fact, constitute a hazard in space engineering.\cite{NASA,Land,NASA_1,ESA,Trib} 
The range of velocities of fast macroscopic projectiles to consider goes 
from a variable and medium-dependent lower limit, 
which will be determined,
to a less variable upper limit, about 10~km/s, which is 
determined, in essence, by basic atomic physics.


The study of the penetration of fast subatomic particles in matter is relatively recent, of course.
Bohr devised an essentially correct theory in 1913 and completed it along 
the following years.\cite{Bohr}
The research on the many aspects of particle penetration has 
played an important role in the development of modern physics and the theory is now well established.\cite{Sigmund}
It is studied in various contexts, from fundamental
particle physics to areas of applied physics, such as 
nuclear engineering or medicine, solid state physics,
etc. In particular, high-energy subatomic particles are part of the space environment, as well as fast meteoroids.\cite{Trib}

Between subatomic and macroscopic projectiles, 
there is a {\em mesoscopic} range of {\em nano-projectiles}, 
with important technological applications.\cite{ion-solid}
The few studies of their relation to macroscopic projectiles\cite{Samela,Anders-etal_PRL} only treat particular aspects of the problem.
Here we study the problem of resistance to projectile penetration within a 
unified conceptual framework that applies to the full ranges of sizes and velocities, and we deduce some novel and useful facts, especially applicable to nano-projectiles.

For the sake of simplicity, we disregard 
the effects of the impact of the projectile on the surface of the medium.
If the penetration depth is considerable, the surface effects (spalling, sputtering, etc) are not significant. 
Another complication is the possible deformation or fragmentation of the projectile,  
considered in Sect.~\ref{hyperv}. These effects reduce the penetration depth. 
As we will see, a fast projectile penetrates deeply when 
the density and strength of the projectile are considerably higher than those of the medium. This condition is less strict for streamlined projectiles.
Subatomic particles are necessarily very fast  
(with velocities larger than 2000 km/s)
and also are very penetrating (for their size).

We begin by setting the basic framework for projectile penetration in Sect.~\ref{fast},
including a dimensional analysis, which allows us to connect with Gamow's formula for penetration of a fast projectile. We need to
define ``fast'' by introducing a critical velocity. 
Next, we analyze the physics of resistance (Sect.~\ref{drag}).
The analysis of resistance to supersonic projectiles
leads us to consider complex high temperature and high pressure phenomena. 
Next, in Sect.~\ref{charged}, the penetration of charged subatomic particles in matter is treated in analogy with that of macroscopic projectiles, beginning by a dimensional analysis. 
A deeper analysis of the electrodynamic origin
of resistance follows (Sect.~\ref{model}).
Finally, we study nano-projectiles, as the 
nexus of the macroscopic and atomic theories of penetration (Sect.~\ref{nano}).

Our subject encompasses two fields of 
physics that are studied by different communities and requires concepts of both fields, some of which are well known by the corresponding community but probably not by the other. For this reason, it is 
preferable to refer to basic articles and textbooks, when possible.
An apology is due to expert readers, who may find some concepts 
and references too basic.

A note on notation. We employ often two signs 
for the asymptotic equivalence of functions: $f \sim g$
means that the limit of $f(x)/g(x)$ is finite and
non-vanishing when $x$ tends to some value (or to infinity),
whereas $f \approx g$ means, in addition, that the limit is one.  
We also use the signs $\propto$ and $\simeq$, which denote, respectively, 
numerical proportionality and approximate equality.
Regarding our notation for functions, we generically employ $f$ to 
denote any one of the many unspecified functions of non-dimensional arguments that we introduce. 

\section{Motion of fast projectiles in a medium}
\label{fast}

The penetration depth, which is easily measurable, can be obtained by 
integrating Newton's Second Law. 
We assume that the motion is one-dimensional, 
as if the projectile is axisymmetric and moves along its axis (the $x$-axis).
Given that the resistance force $F$ is a function of the velocity $v$, 
the differential equation of motion can be solved for $t(v)$
and hence $v(t)$ can be obtained.
One more integration gives 
$x(t)$. 
However, to obtain the
penetration depth, $t$ is not a relevant variable and, instead, 
we are interested in the function
$v(x)$. 
Therefore, we write 
\begin{equation}
M v\,\frac{dv}{dx} = -F(v),
\label{x-v}
\end{equation}
where $M$ is the mass of the projectile.
Integrating $dx$
between $x(v_0)=0$ and $x(0)=D$
(the penetration depth) one gets:
\begin{equation}
\frac{D}{M} = \int_0^{v_0} \frac{v\,dv}{F(v)}\,.
\label{D-int}
\end{equation}
Naturally, Eq.~(\ref{x-v}) is equivalent to the kinetic energy theorem
and the braking can be understood alternately as a loss of
momentum or as a loss of kinetic energy.

The resistance force depends
on the geometry of the body but is independent of its mass, as long as
gravitation plays no role in the dynamics of medium resistance.
Therefore,
Eq.~(\ref{D-int}) implies that the penetration depth, $D$, is proportional 
to $M$.  

\subsection{A fundamental law of motion under resistance}
\label{fund}

The force of resistance 
is, basically, a certain combination of the Coulomb electric forces of interaction between the microscopic particles that constitute the body and those that constitute the medium,
namely, electrons and atomic nuclei. 
The (very large) set of equations of motion of
these particles possesses an invariance: the
multiplication of {\em all} the present masses by a common factor $\l$ can be
compensated by the multiplication of time by $\l^{1/2},$ while the space coordinates and electric charges are kept constant.
This operation does not change the trajectories and, hence,
it does not change $D$,
provided that the initial velocity $v_0$ is scaled to $\l^{-1/2}v_0$. 
Therefore, we deduce that $D$ is proportional to $Mv_0^2$ times some 
function of $mv_0^2$, where $m$ is the mass of the particles of the medium 
that interact with the body
(if there are several types of them, then the corresponding terms appear). 
Furthermore, we deduce from Eq.~(\ref{D-int})
that $v$ must appear in $F(v)$ as the combination $m^{1/2}v.$
In a continuous medium, described by the mass density $\rho_m$,
$m^{1/2}v$ is replaced by $\rho_m^{1/2}v$. 

This fundamental law is upheld by the concrete dimensional analyses for 
macroscopic bodies in the next section (Sect.~\ref{dim_anal}) and for 
subatomic particles (Sect.~\ref{dim_anal_1}) and, furthermore, is generally 
useful (e.g., in Sect.~\ref{low-v}).

\subsection{Macroscopic body penetration: dimensional analysis}
\label{dim_anal}

The minimal set of parameters for penetration in a continuous medium
consists of 
the impact velocity $v_0$, mass $M$ and length $l$ of the body, its
set of non-dimensional shape parameters $\{s\}$, 
and the density $\rho_{\mathrm{m}}$ of the medium. 
The penetration depth $D$ must be a function of these magnitudes.  As seen above, $D$ is proportional to $M$.
Therefore, we define $\tilde{D} = D/M$ 
and we have to find the function $\tilde{D}(v_0, l, \{s\}, \rho_{\mathrm{m}})$.  
We first notice that, of the base mechanical dimensions L, T and M, the dimension T is only present in 
$v_0$. Therefore, the dimension T cannot be present in a
non-dimensional group and, consequently, $\tilde{D}$ cannot depend on $v_0$. 
Since the shape parameters are non-dimensional, the dimensions of $\tilde{D},$
namely, $[\tilde{D}] = {\rm L} {\rm M}^{-1},$ must be obtained with a
combination of $l$ and $\rho_{\mathrm{m}}\,.$ The only possible
combination is ${1}/(\rho_{\mathrm{m}}\,l^2).$
In consequence,
\begin{equation}
\tilde{D}(v_0, l, \{s\}, \rho_{\mathrm{m}}) = 
\frac{f(\{s\})}{\rho_{\mathrm{m}}\,l^2}\,. 
\label{D-Gs}
\end{equation}
where $f$ is a non-dimensional function of the non-dimensional shape parameters.

Let us compare this formula with empirical scaling formulas, which adopt the form:\cite{Kinslow,Ben,Chinamen,NASA,Land,Gold}
\begin{equation}
D = k \,\rho_M^\a\, \rho_{\mathrm{m}}^{\a'} v_0^\b\, l^\g,
\label{D-empirical}
\end{equation}
where $\rho_M \propto M/l^3$ is the density of the body,
$k$ is a dimensional constant, and 
the exponents $\a$, $\a'$ and $\b$ have been assigned various values.  
Since $D$ has to be proportional to $M$, 
$\a=1$ (but other values appear in the literature).
It is often assumed that $\a+\a'=0,$ that is to say, that $D$ depends on
$\rho_M$ and $\rho_{\mathrm{m}}$ through their ratio,
like in Eq.~(\ref{D-Gs}).
The exponent $\b$ is usually smaller than one (a typical
value is 2/3), so that $D$ has a weak dependence on $v_0$.
However, $\a$ usually is smaller than $\b$, which implies that the dependence 
of $D$ on the density ratio is still weaker. Naturally, Eq.~(\ref{D-Gs}) 
indicates the opposite.

Eq.~(\ref{D-Gs}) is the simplest scaling formula and it agrees with Gamow's 
formula,\cite{Gamow}
\begin{equation}
D = \frac{\rho_M}{\rho_{\mathrm{m}}}\,l,
\label{D-G}
\end{equation}
except for the shape-dependent numerical factor.
Of course, Gamow must have been aware of the possible presence of this factor,
because he loosely defined $l$ as the ``length of the projectile,'' 
without specifying any shape, and commented that Eq.~(\ref{D-G}) 
``is true only very approximately.''  
Gamow's formula is useful for estimations. For example, a fast lead bullet 
will penetrate 11 times its length in water (or flesh) and 
$11\hspace{1pt}000$ times in air.
It also provides the rationale 
for high density projectiles as armor penetrators, as well as for high 
density armor plates, which favors the use of depleted 
uranium.\cite{DU}
Gamow remarked that ``the length of penetration does not depend on the initial velocity of the projectile (provided that this velocity is sufficiently high).'' 
To justify it, he mentioned some experimental evidence, without references.
The absence of $v_0$ in Eq.~(\ref{D-Gs}) 
is, in fact, a consequence of our choice of the minimal set of parameters, 
in which only $v_0$ 
includes the time dimension. In this regard, one may question the meaning of
``sufficiently high'' initial velocity, as long as there is no velocity
with which to compare it.

\subsection{Critical velocity}
\label{v_c}

A critical velocity is necessary to qualify a projectile as ``fast.''
The speed of sound, $c_{\mathrm{s}}$, is the characteristic velocity of any medium; its fundamental role
in penetration dynamics is shown in Sect.~\ref{physics}. 
But it is not the only option:
in incompressible fluids, with $c_{\mathrm{s}}=\infty$, 
the critical velocity is given by viscosity.\cite{Saslow} 
Regardless its origin, let us denote the critical velocity by $v_{\mathrm{c}}$
and just add it to the minimal set of parameters. 
Then, $v_0$ and $v_{\mathrm{c}}$ must appear in $\tilde{D}$
as the ratio $v_0/v_{\mathrm{c}}\,.$ Therefore, Eq.~(\ref{D-Gs}) is
replaced by
\begin{equation}
\tilde{D}(v_0/v_{\mathrm{c}}, l, \{s\}, \rho_{\mathrm{m}}) = 
\frac{f(\{s\}, v_0/v_{\mathrm{c}})}{\rho_{\mathrm{m}}\,l^2}\,. 
\label{D-vc}
\end{equation}

This formula is very general yet simpler than others that 
have been proposed for particular cases.
For example, let us compare it to Li and Chen's formula for penetration in concrete.\cite{Li-Chen}
Li and Chen express $D$ in terms of a non-dimensional function of
three non-dimensional variables, which we can choose as: (i)~the {\em nose 
factor} (one shape parameter);
(ii)~$v_0/v_{\mathrm{c}}$, where $v_{\mathrm{c}}$ is expressed in terms of 
target material properties;
and (iii)~$M/(\rho_
{\mathrm{m}}l^3)$. If we take into account that 
the dependence on this third variable is 
actually determined by basic physical principles, as already said,
Li and Chen's formula boils down to a particular case of Eq.~(\ref{D-vc}).

Eq.~(\ref{D-vc}) reduces to Eq.~(\ref{D-Gs}) when $v_0/v_{\mathrm{c}} \ra \infty$,
provided that $f(\,\cdot\,,v_0/v_{\mathrm{c}})$ has a finite,
non-zero limit. 
When a non-dimensional function has a finite, non-zero limit, the asymptotics is said to be of the first kind.\cite{Baren} 
The next more complex case is 
{\em asymptotic similarity}, namely, power-law behavior.\cite{Baren} 
A logarithmic asymptotic law is a border-line case, because the logarithmic law can be understood as a power law of zero exponent.  
In fact, the asymptotics that corresponds to
Newton's theory is logarithmic (Sect.~\ref{empiric}). 
This means that the penetration depth has a very weak
dependence on the critical velocity, so that this velocity defines the
asymptotic regime and {\em almost} disappears in it.

\section{Models of resistance}
\label{drag}

\subsection{Empirical models}
\label{empiric}

$D(v_0)$ can be measured and hence the function $F(v)$ can be determined empirically.
Newton\cite{Newt} already conducted experiments on air drag and proposed the
empirical law $F(v) = A v + B v^{3/2} + C v^2.$ 
Newton's own measurements imply that the middle term is negligible in
comparison with the other two terms.\cite{Saslow} 
The quadratic polynomial $F(v) = a_1 v + a_2 v^2$ has 
been much employed.\cite{Resal,Saslow,Gold,Ben} 
Substituting this $F(v)$ in Eq.~(\ref{D-int}), 
\begin{equation}
\tilde{D} = \int_0^{v_0} \frac{dv}{a_1 + a_2\, v} = 
\frac{1}{a_2}\,
\ln \left(1 +  \frac{a_2\,v_0}{a_1} \right).
\label{D-SL}
\end{equation}
Comparing this equation with Eq.~(\ref{D-vc}),
we identify $a_2 \propto
\rho_{\mathrm{m}}l^2$ and $v_{\mathrm{c}} = a_1/a_2\,.$ This $v_{\mathrm{c}}$ 
is the velocity at which the linear and the quadratic terms 
of $F$ have equal magnitude.
For $v \gg v_{\mathrm{c}},$ we can just use $F(v) \approx a_2 v^2,$ 
but the corresponding integral, $\int_0^{v_0} dv/v$,
diverges at its lower limit, because a purely
quadratic drag is too weak at low velocity 
to stop the projectile. 
Naturally, 
the lower limit must be set to $v_{\mathrm{c}}$, obtaining
\begin{equation}
\tilde{D} \approx \frac{1}{a_2}\,\int_{v_{\mathrm{c}}}^{v_0} \frac{dv}{v}
= \frac{1}{a_2}\,\ln \frac{v_0}{v_{\mathrm{c}}}
\propto \frac{1}{\rho_{\mathrm{m}}\,l^2}\,\ln \frac{v_0}{v_{\mathrm{c}}}
\,.
\label{D-q}
\end{equation}
This form is asymptotically equivalent to the exact result (\ref{D-SL}) 
but is more general, because it holds whenever $F(v) \sim v^2$ 
for $v \gg v_{\mathrm{c}}$.
On the other hand, if 
$F(v) \approx a_1 v$, then $\tilde{D} \approx v_0/a_1,$ which agrees with 
the asymptotic limit of (\ref{D-SL})
for $v \ll v_{\mathrm{c}}.$ 

Recalling the empirical power-law form of $D(v_0)$ in 
Eq.~(\ref{D-empirical}), we can propose $F = a v^{2-\b}$, which gives $\tilde{D} = v_0^{\b}/(\b a)$. To have $0 <\b <1$, as usually found, the exponent of $v$ in $F$ must be between 1 and 2. This 
contradicts the conclusions drawn from Newton's experiments.
However, it is the expected result for 
a function such as the one in Eq.~(\ref{D-SL}) that
one tries to fit to a power-law in 
a non-asymptotic range 
(as further discussed in Sect.~\ref{physics}).

For penetration in solids at low-velocity, the force is actually
constant\cite{Robins,Euler,Poncelet,Ben,Li-Chen} 
and $\tilde{D} \sim v_0^2.$
The full quadratic polynomial, $F(v) = a_0 + a_1 v + a_2 v^2$, gives rise to an integral $\int v\,dv/F$ that can be effected
by factorizing the polynomial, but
the result is a complicated algebraic formula 
in terms of $a_0$, $a_1$ and $a_2$.
Nevertheless, the asymptotic form (\ref{D-q}) holds. Now, 
$v_{\mathrm{c}}$ is to be identified with $a_1/a_2$,
if $a_1^2 > 4a_0a_2$, or with $(a_0/a_2)^{1/2}$, if $a_1^2 \leq 4a_0a_2$. 

We can try a higher power of $v$ to modify the asymptotics; for example,
$F(v) = a_1 v + a_2 v^2 + a_3 v^3$
($a_1,a_2, a_3 >0$). 
Then we have two critical velocities:
$v_{\mathrm{c}} = a_1/a_2$ and $v_{\mathrm{c}}' = a_2/a_3\,.$
The integral for $\tilde{D}$ yields a complicated function, which, remarkably,
has a large-$v_0$ asymptotics of the first kind. In particular, if
$v_{\mathrm{c}}/v_{\mathrm{c}}' = a_1 a_3/a_2^2 \ll 1,$ then
$
\tilde{D} \approx \ln (v_{\mathrm{c}}'/v_{\mathrm{c}})/a_2.
$
Comparing it to (\ref{D-q}), we see that
$\ln(v_0/v_{\mathrm{c}})$ is replaced by
$\ln(v_{\mathrm{c}}'/v_{\mathrm{c}}),$ as if the largest possible value of $v_0$ were $v_{\mathrm{c}}',$ namely, the velocity for equality of the quadratic and cubic terms. 
Actually, the effect of the cubic term on a motion with 
$v_0 > v_{\mathrm{c}}'$
is to bring $v$ down to 
$v_{\mathrm{c}}'$ while a distance $\approx 1/a_2$ is traversed,
as easily found by making $F(v) \approx a_3 v^3$.
The subsequent motion is due to the quadratic drag, $F(v) \approx a_2 v^2$, and traverses a longer distance $\approx \ln(v_{\mathrm{c}}'/v_{\mathrm{c}})/a_2$. Finally, under $F(v) \approx a_1 v$, the distance traversed is $\approx 1/a_2$.

Generalizing these results, we conclude that 
the quadratic force, $F(v) = a_2 v^2$, is singled out because the integral 
$\int_0^\infty v\,dv/F$ is divergent at both the lower and upper limits. 
When these divergences are cut off, 
the factor $\ln(v_{\mathrm{c}}'/v_{\mathrm{c}})$ arises.
Any addition to the quadratic force that is stronger than it at high velocities makes the integral converge at $v=\infty$ and works like
the cubic term, by slowing down the projectile to the corresponding 
$v_{\mathrm{c}}'$, within a distance $\sim 1/a_2$.
Analogously, any addition that is stronger than it at low velocities makes the integral converge at $v=0$ and, like the linear term, stops the projectile from $v_{\mathrm{c}}$, within a distance $\sim 1/a_2$.
Moreover, the quadratic force is singled out 
because $a_2$ does not involve the dimension T and can be 
expressed in terms of the basic variables, namely, 
$a_2 \propto \rho_{\mathrm{m}}l^2$.

Finally, let us notice that the coefficients $a_0$, $a_1$ and $a_2$ 
are necessarily positive, but we can have
$a_3 < 0$ or, in general, that the growth of $F$ with velocity is lower than quadratic for very large velocities (Sects.~\ref{model} and \ref{nano}). 
Some recent experimental results about the relevant velocity ranges
are cited in Sects.~\ref{physics}, \ref{hyperv} and \ref{nano}.

\subsection{Physical origin of the force of resistance}
\label{physics}

Newton's argument is, in essence, very simple: 
the resistance force results from the total backward force of
reaction of the particles of the medium to 
the forward forces exerted on them by the moving body.
In modern terms, the force
is the consequence of momentum conservation in the continuous collision of the
body with the particles of the medium. 
Newton actually considered a specific model that yields the force dependence on the shape of the body, and, remarkably, he calculated with that model the profile of revolution of least resistance.\cite{N-ex} 

Newton's theory can be put in a general form that is independent of 
the type of body-medium interaction and is applicable to subatomic particles
(Sect.~\ref{LL-J_calc}).
As the body moves at velocity $v$, 
it crosses, in a time $d t$, a slab of the medium of depth $v\,d t$. 
This slab is formed by mass elements 
$dm = \rho_{\mathrm{m}} \,dS\, (v\,dt)$, where $dS$ is the element of surface in a transversal plane. 
If the body imparts to an impinging mass element $dm$ a longitudinal velocity 
$v_{\parallel}$, then it transfers to it a longitudinal momentum 
$dp = \rho_{\mathrm{m}} \, dS \,(v\,dt)\,v_{\parallel}$ and exerts a force
$dF = dp/dt = \rho_{\mathrm{m}} v\,v_{\parallel}\, dS.$ Therefore, 
the total drag force is
\begin{equation}
F = \rho_{\mathrm{m}}v \int \!v_{\parallel}\, dS.
\label{F-Newt}
\end{equation}
The mass elements may also acquire transversal velocity $v_{\perp}$,
but this is irrelevant to $F$.

Newton {\em assumed} that $v_{\parallel}$ is proportional to $v$. 
This condition seems 
natural but is not universal (see Sects.~\ref{LL-J_calc} and \ref{nano}). 
If we write 
$v_{\parallel} = c\,v,$ where $c$ is a geometrical coefficient, 
then 
\begin{equation}
F = \rho_{\mathrm{m}}v^2 \int \!c\, dS.
\label{F-Newt_1}
\end{equation}
In particular, if
the mass elements placed inside the cylinder defined by the cross section of 
the body undergo completely inelastic collisions with the body, without adhering to it,
while the other mass elements are unaffected, then $c=1$ in the cross section
and zero out of it. Therefore,  
$F = \rho_{\mathrm{m}}v^2 A$, where $A$ is the cross-sectional area.
In Newton's specific model, the collisions are inelastic only in the direction normal to the surface of the body.
In general, the collisions are partially elastic and, furthermore, the body can affect mass elements close to its cross section. 
Therefore, the integral in  
Eq.~(\ref{F-Newt_1}) defines an effective cross-sectional area, which depends on the shape of the body and its interaction with the medium. 
In fluid mechanics (see below), the force is normally
written as $F = \rho_{\mathrm{m}}v^2 (C_{\mathrm{D}}/2)A$, including a \emph{drag coefficient} 
$C_{\mathrm{D}}$ that depends on the shape of the body and 
is of the order of unity 
for bluff bodies but much smaller for streamlined bodies.\cite{L-L_FM}

Following Newton,
Gamow\cite{Gamow} argued that the projectile pushes aside the medium, boring a tunnel, and that
``it is easy to see that the sidewise velocity of the medium 
is about the same as the velocity of the advancing projectile,''
that is to say, that $c=1$ in the cross section.
However, Gamow's conclusion, namely, ``the
projectile will stop when the mass of the medium moved aside is of the same
order of magnitude as its own mass," is a {\em non sequitur}, because 
it is necessary to solve the differential equation of the motion.\cite{Saslow}
As seen in~(\ref{D-q}), the integration demands the cutoff $v_{\mathrm{c}}$ and produces the factor 
$\log(v/v_{\mathrm{c}})$. 
To be precise, employing fluid mechanics notation, namely, 
$a_2 = \rho_{\mathrm{m}}(C_{\mathrm{D}}/2)A$, one obtains
\begin{equation}
D = \frac{M}{\rho_{\mathrm{m}}(C_{\mathrm{D}}/2)A}\,\ln \frac{v_0}{v_{\mathrm{c}}} = \frac{\rho_M\,l}{\rho_{\mathrm{m}}(C_{\mathrm{D}}/2)}\,\ln \frac{v_0}{v_{\mathrm{c}}}
\,,
\label{D-CD}
\end{equation}
where, for the second equality, 
the volume of the projectile has been set to $lA$ (which is exact for a cylinder).
Both the two factors that correct Gamow's formula, namely,
the logarithm and $(C_{\mathrm{D}}/2)^{-1}$, make $D$ grow, 
and the latter can easily be larger than the former. Therefore, a streamlined shape is more important than a high velocity for a projectile to be penetrating, and streamlined projectiles can widely surpass Gamow's $D$.
However, the asymptotic form (\ref{D-CD}) is then less applicable, 
because $v_{\mathrm{c}}$ increases as $a_2$ decreases. The reason is that 
$v_{\mathrm{c}}$ is defined as the value of $v$ such that $a_2 v^2$ equals 
the remainder of $F(v)$, which is less sensitive to shape (see below).

In incompressible fluids, the origin of drag is viscosity, namely, 
internal friction or resistance to the relative motion 
of contiguous fluid layers.
For a fluid characterized by its density $\rho_{\mathrm{m}}$ and viscosity coefficient $\eta$, 
the dimensional analysis requires that
$F = \rho_{\mathrm{m}}\, l^2 v^2 f(\{s\},\mathrm{Re})$, 
where $f$ is a non-dimensional function and 
$\mathrm{Re} = v\,l \rho_{\mathrm{m}}/\eta$ is the Reynolds number.\cite{Baren,L-L_FM} 
The force is best written as
$$
F = \rho_{\mathrm{m}}\, A \,v^2\, C_{\mathrm{D}}(\{s\},\mathrm{Re})/2,
$$
with a velocity-dependent drag coefficient. This coefficient 
must have a limit when $\mathrm{Re} \ra \infty$,
so that $\eta$ disappears from $F$, 
as corresponds to inertial motion. 
One could think that there should be no friction when the viscosity vanishes, but the transfer of momentum does not vanish: it concentrates on a {\em 
boundary layer} of vanishing width around the body that gives rise to a fluid
{\em wake}. 
For slow body motion, namely, for $\mathrm{Re} \ll 1$, the inertia of 
the body is negligible in comparison with the effect of viscosity, so 
that $\rho_{\mathrm{m}}$ disappears from $F$. This implies that 
$C_{\mathrm{D}} \sim 1/\mathrm{Re}$ and
$$
F = \eta\, l v f(\{s\}),
$$
which is the general form of the Stokes drag law. In this case, $F$ is not very sensitive to shape, so $f(\{s\}) \gtrsim 1$ 
even when $A \ll l^2$ (being $l$ the length of the body).\cite{Loth}

In compressible fluids, 
we have an additional parameter, the compressibility coefficient, or its 
inverse, the bulk modulus $K$. We neglect
viscosity to simplify the analysis and characterize the fluid just by 
$\rho_{\mathrm{m}}$ and $K$.
Given that $[K] = {\rm L}^{-1}\,{\rm T}^{-2}\,{\rm M},$
we must have 
$F(v) = \rho_{\mathrm{m}}\, l^2 v^2 f(\{s\},v/c_{\mathrm{s}})$,
where $c_{\mathrm{s}} = (K/\rho_{\mathrm{m}})^{1/2}$ is the velocity of sound
and plays the role of critical velocity. 
When $v \gg c_{\mathrm{s}}$ (hypersonic motion), the resistance is inertial and we can write $F = (C_{\mathrm{D}}/2)\rho_{\mathrm{m}} A\, v^2$.
At low velocity, we should obtain a $v$-independent force, namely,
$F = f(\{s\}) Kl^2$, which would result from  
the balance of the fluid pressures around the body. However, 
the fluid is incompressible at low $v$ and this force is null.\cite{L-L_FM}

Let us compare the critical velocities in incompressible and compressible fluids, which are,
respectively, $\eta/(l \rho_{\mathrm{m}})$ and $c_{\mathrm{s}}$ (for bodies of generic shape). The latter is a characteristic
of the fluid whereas the former also depends on the length of the body. 
The value of $l$ such that they are equal,
namely, $\eta/(c_{\mathrm{s}}\,\rho_{\mathrm{m}})$, 
is a characteristic of the fluid: 
it is, approximately, the molecular {\em mean free path} $\l$, which is an 
important length in molecular kinetics.\cite{Reif}
It is a microscopic length in standard conditions 
and then $c_{\mathrm{s}}$ is the larger critical velocity.
However, molecular kinetics is relevant to the motion of macroscopic projectiles in rarefied gases and to the motion of nano-projectiles (Sect.~\ref{nano}). 
Normal macroscopic projectiles, with $l \gg \l$, can undergo subsonic or supersonic inertial resistance,
the former due to viscous transfer of momentum 
and the latter due to compressional processes that raise 
the value of $C_{\mathrm{D}}$.

For penetration in solid media, the only critical velocity
is the speed of sound, which is indeed the standard reference velocity for 
fast projectile impacts onto solid targets. 
The resistance force to a hypersonic projectile must be, like in fluids, 
$F = a_2 v^2$, where $a_2 \propto \rho_{\mathrm{m}}\,A$ and is 
shape dependent. At low velocity, 
a constant force results from tangential surface-friction forces.
Therefore, the force is normally modeled as $F = a_0 + a_2 v^2$,
although it can also have a linear term.\cite{Poncelet,Ben,Li-Chen}
A popular model of cylindrical projectile penetration sets
$a_2=N^* B \rho_{\mathrm{m}} A$,
where the ``nose factor'' $N^*$ is the only shape parameter and $B \simeq 1$ is an extra parameter of the target material.\cite{Li-Chen}  

In solids as well as in fluids, a supersonic body generates a {\em bow wave}, which is a type of shock wave (discontinuity of the velocity field).
In a bow wave, there is a sharp rise of pressure at the bow of the body
that weakens far from it and becomes an ordinary sound wave.\cite{L-L_FM}
The wave carries away momentum of the body and generates resistance 
to its motion. 
A hypersonic bow wave front lies very close to the forward surface of 
the body, where the velocity field has a sharp discontinuity, such that
the streamlines are bent almost tangentially to the surface, 
like in Newton's specific model of resistance.\cite{Ander,Zel}
The general characteristics of a
shock wave discontinuity are given by the Hugoniot relation, 
in terms of the thermodynamic properties of the medium.\cite{L-L_FM,Zel}
As regards these properties, we can distinguish gases, in which the pressure has thermal origin, from condensed media, in which its origin is the short-range atomic repulsion.
Gases are usually considered to be {\em polytropic}, with bulk modulus proportional to pressure, namely, $K = \g P$, 
where the index $\g$ is characteristic of the gas ($1<\g\leq 5/3$).
In condensed media, 
$K$ is large and is non null for vanishing $P$.
The equation of state of solids is a subject of study.\cite{Kinslow,Zel} 
Some basic shock-wave physics is introduced in Sect.~\ref{hyperv}. 

As already noticed, the inertial resistance to penetration 
can be greatly reduced by shape optimization, with the effect of raising 
the value of the critical velocity. 
In condensed media and, in particular, in solids, with large values of $c_{\mathrm{s}}$, 
the critical velocities of optimized projectiles can be exceedingly large.
For example, tests of the cylindrical projectile model with 
$a_2 = N^* B \rho_{\mathrm{m}} A$  against experiments in which $N^* < 0.2$ 
and $B=1.0$ find a rather strong dependence of $D$ on $v_0$,\cite{Li-Chen} 
while the asymptotic law (\ref{D-CD}) (with $C_{\mathrm{D}}/2=N^* B$) predicts a weak dependence.
The reason is that $N^*$ is so low and hence $v_{\mathrm{c}}$ so large
that the asymptotic regime is not approached even at the highest velocities tested. 
This situation is surely common and suggests that the empirical law 
$D \sim v_0^{\b}$ and the corresponding force $F \sim v^{2-\b}$, 
with $\b \simeq 2/3$, result from numerical fits
in velocity ranges below the asymptotic inertial regime. 

\subsection{The fate of a hypersonic projectile}
\label{hyperv}

The Hugoniot relation implies that mechanical energy is dissipated in a shock wave, that is to say, is {\em irreversibly} lost.
The ultimate reason for this irreversibility is the loss of {\em adiabaticity} at the shock front, 
because the medium cannot adjust to the rapid mechanical change (adiabatic invariance is studied in mechanics\cite{L-L_M}).
As the medium is set in motion, it gains kinetic and internal energy,\cite{shock-dissip}
but the dissipation is measured by the increase of entropy.
Its magnitude allows one to classify shock waves as weak or strong.\cite{L-L_FM,Zel} 
Weak shock waves involve small compressions and are close to sound waves.
Hypersonic penetration generates strong shock waves, 
in principle. However, in the case of 
a slender projectile that disturbs the medium only slightly, 
the bow wave is strong only in the very hypersonic limit.

The rises of temperature and pressure in a bow wave   
can be high enough to alter the projectile. In gases, 
the main effects are thermal.\cite{Ander,Zel}
For example, it is known that a large fraction of the mass of a meteorite 
is lost by ablation while traversing the atmosphere.
A simple model of ablation\cite{Bronshten} predicts that the fraction of 
the mass of a meteorite lost while braking in air 
only depends on the initial velocity and is $1-\exp(-\s v_0^2/2),$
where $\s \simeq 10^{-8}$ kg/J. Therefore, 
the mass loss is considerable for $v_0=10$~km/s but negligible for $v_0=1$~km/s. 
 
In a condensed medium, even a weak shock wave can generate high pressures,
in spite of being almost adiabatic and isothermal. The process is 
ruled by the ``cold'' equation of state of the medium, which can be derived from the intermolecular potential through the virial theorem. For example, the power-law potential 
$V(r) \sim r^{-s}$ (e.g., the repulsive part of the Lennard-Jones potential) gives $P/P_0 = (\rho/\rho_0)^{s/3+1}$ and $K = (s/3+1)P$. 
This is a polytropic equation with a large index 
(equal to 5 for $s=12$).
Naturally, the attractive part of the potential is necessary to have non-null $K_0 = K|_{P=0}$. 
A suitable law is $K = K_0 + nP$, in which $n$ empirically goes from 4, 
for metals, to 7, for water, and $K_0$ is given by the standard speed of 
sound ($c_{\mathrm{s}}= 1.5$ km/s and $K_0 = 2.2$ GPa for water, and larger 
for solids).\cite{Zel}
For $P \ll K_0/n$, $K$ can be taken constant (Hooke's law). 
Even for $P = K_0/n$, the compression ratio is small
(10\% for $n=7$, e.g., for water), and so is the rise of temperature. 

However, in the absence of notable thermal effects, the stresses can
deform the projectile. Furthermore, its material becomes {\em plastic} 
for stresses much smaller than its own $K_0$ value (assumed to be very large). 
This process can involve fracturing, so
the projectile can break in fragments.\cite{Bronshten,Zukas,Kinslow,Iskander,Ruiz-S,Ben,Chinamen,Kadono}
The analysis of deep penetration into low-density materials 
agrees with the law~(\ref{D-CD})
for impact velocities below some critical velocity for fragmentation 
(a few km/s).\cite{Kadono}
Naturally, projectiles with optimal shapes undergo much smaller stresses
and their critical velocities for fragmentation are higher.

Condensed media are strongly heated by strong shock waves and become, 
under the corresponding extreme pressures, high-density gases.
After a hypervelocity impact, the medium at the front of the projectile is compressed, heated and, if solid, fluidified, 
while it expands and cools at its back.
Simultaneously, a shock wave propagates through the body backwards, until it reaches the end and unloads. 
If the specific kinetic energy of the body is about ten times 
the heat of vaporization of its material, then
the unloading of the shock wave releases as much energy as to completely 
vaporize the body.\cite{Zel}
This happens at impact velocities over 10 km/s. In general, 
a fast projectile must be considered as a bunch of individual atoms when  
its kinetic energy is much greater than the cohesive energy of those atoms. 
Heats of vaporization for solids are of several eV per atom, 
equivalent to velocities of several km/s [1~eV/atom = $9.65\cdot 10^4$ J/mol,
J/kg = (m/s)$^2$].
Besides, hypersonic penetration can 
induce internal changes in molecules or atoms, such as molecular excitation 
or dissociation, chemical reactions and even ionization.\cite{Ander,Zel}
At much larger velocities, we can neglect the binding of electrons to atomic nuclei and consider the penetration of high-energy charged particles.

\section{Motion of fast charged particles in matter}
\label{charged}

The crucial difference between the interactions of macroscopic bodies
and of charged subatomic particles while moving through matter is
the difference in interaction range, which
combines with the difference in size.
Actually, the size of a material body is defined by the relative positions of its neutral atoms or molecules, which are kept in place by  
low-energy and short-range electric
interactions with the nearby atoms or molecules 
similar to the interactions that they have with the particles of the medium, 
whereas, at high energy, the charged subatomic constituents
interact directly through the long-range Coulomb force. 

Therefore, the basic parameters change: 
the body size $l$ is substituted by the charge $Q$, while the other two parameters of the object, namely,
$M$ and $v_0$ stay the same. Besides, we need
some electrical magnitude to characterize the medium. 
In an electrically neutral medium, 
a charged moving particle can only induce a relative 
displacement of negative and positive charges, namely, it must induce 
currents or polarization.
In fact, the displaced charges are, mostly, mobile electrons. 
However, we do not have to specify the nature of the 
mobile particles yet and we
treat them as constituting a \emph{charged continuum} defined only by its charge density, denoted by $\rho_e$. 
Naturally, $e$ will eventually refer to the electron charge. 
The mechanical parameter of the medium, namely, its mass density, is now ambiguous, because it may refer to the total mass density of the neutral medium or just to the mass density of the particles 
associated with the charge density $\rho_e$. 
We denote it by $\rho_m$, meaning the mass density of particles of mass $m$, without specifying their nature yet.

In summary, we assume in our modeling that the characteristic magnitudes are:
$M$, $Q$, and $v_0$, belonging to the moving particle, plus $\rho_m$ and
$\rho_e$, belonging to the medium. All are positive, by default.
We have one more magnitude than in the case of macroscopic bodies. 

\subsection{Dimensional analysis}
\label{dim_anal_1}

We redefine the electric charge to absorb the factor $4\pi \e_0$, namely,
$q/\sqrt{4\pi \e_0} \ra q$, as in the Gaussian or electrostatic unit systems, so that $[q] = {\rm L}^{3/2}\,{\rm T}^{-1} {\rm M}^{1/2}$
and we only have mechanical dimensions.
To find $\tilde{D}(Q, v_0,  \rho_e, \rho_m)$,
we employ the standard method. 
Let us write 
\begin{equation}
\tilde{D}(Q, v_0,  \rho_e, \rho_m) = Q^\a\,v_0^\b\, \rho_e^\g\, \rho_m^\d\,,
\label{D-powerlaw}
\end{equation}
with unknown exponents. From $[\tilde{D}] = {\rm L} {\rm M}^{-1},$ we obtain:
\begin{eqnarray}
{\rm L}&:& \phantom{-}1 =  \frac{3\a}{2} + \b - \frac{3\g}{2} - 3\d;
\label{B1}
\\
{\rm T}&:& \phantom{-}0 =  -\a - \b - \g;
\label{B2}
\\
{\rm M}&:&           -1 =  \frac{\a}{2} + \frac{\g}{2} + \d.
\label{B3}
\end{eqnarray}
This is a linear system of 3 equations with 4 unknowns. 
The matrix of coefficients
has maximal rank, as proved by computing the determinant of the first
$3 \times 3$ matrix, corresponding to $\a,\;\b$ and $\g$. Therefore, the system
is soluble and has an infinite number of solutions, given by a particular
solution plus the general solution of the homogeneous system. Since 
the matrix for $\a,\;\b$ and $\g$ is non-singular, 
we solve the homogeneous system for them, finding
$\a = -2\d/3, \; \b = 2\d, \; \g = -4\d/3.$ 
Therefore, the non-dimensional parameter (pi-group) is
\begin{equation}
\Pi = \frac{v_0^2\,\rho_m}{Q^{2/3}\rho_e^{4/3}}\,.
\label{Pi}
\end{equation}

As a particular solution of system
(\ref{B1},\ref{B2},\ref{B3}),
we prefer the one with $\b=0$, to obtain an expression of 
$\tilde{D}$ that does not contain $v_0$. The corresponding determinant is
non-vanishing and we find:
$\a = -2/3, \; \g = 2/3, \; \d = -1.$
In conclusion,
\begin{equation}
\tilde{D}(Q, v_0,  \rho_e, \rho_m) = 
\frac{\rho_e^{2/3}}{Q^{2/3}\,
  \rho_m}\,f\!\left(\frac{v_0\,\rho_m^{1/2}}{Q^{1/3}\rho_e^{2/3}}\right),
\label{D-f}
\end{equation}
where $f$ is an arbitrary non-dimensional function of its non-dimensional argument (chosen as $\Pi^{1/2}$, to make it proportional to $v_0$).  

\subsection{Physical meaning of the variables and preliminary analysis}
\label{nature}

We have found, in our simple model, a built-in characteristic velocity; namely, 
\begin{equation}
v_{\mathrm{c}} = \frac{Q^{1/3}\rho_e^{2/3}}{\rho_m^{1/2}}\,.
\label{vc}
\end{equation}
Therefore, a first-kind $v_0$ asymptotics of Eq.~(\ref{D-f}) would give
an even simpler model, namely,
$\tilde{D} \sim {\rho_e^{2/3}}/(Q^{2/3} \rho_m).$
This formula is analogous to Eq.~(\ref{D-Gs}),
with the length of the projectile replaced by 
the characteristic length $l_{\mathrm{c}}=(Q/\rho_e)^{1/3}$ (and no shape
parameters). 
$l_{\mathrm{c}}$
is the linear dimension of the volume of the medium that 
contains a total charge equal to $Q$.
If we assume that the density of charge $\rho_e$
corresponds to mobile electrons, then $\rho_e= e\,n$, 
where $e$ and $n$ are the electron charge and
number density, respectively. So we can write 
$$
l_{\mathrm{c}} = 
z^{1/3}\, n^{-1/3},
$$
where $z = Q/e$ and
the length $n^{-1/3}$ is the linear dimension of the volume per
electron or, loosely speaking, the interelectronic distance.  
For a subatomic particle, 
$z \gtrsim 1$, and $l_{\mathrm{c}}$ is not much larger than the interelectronic distance.

We can write the relation $\tilde{D} \sim {\rho_e^{2/3}}/(Q^{2/3} \rho_m)$
in a more convenient form: 
\begin{equation}
\frac{D}{l_{\mathrm{c}}} \sim 
\frac{M}{\rho_m\,l_{\mathrm{c}}^3} =
\frac{e}{Q}\,\frac{M}{m} = \frac{e/m}{Q/M}\,,
\label{D/l}
\end{equation}
where $m= \rho_m\,n^{-1}$ is the mass of the volume that contains one mobile
electron. 
Assuming that $\rho_m$ is just the mass density of 
mobile electrons, $m$ is the electron mass and 
the quotient $D/l_{\mathrm{c}}$ is the quotient of the 
electron charge-to-mass ratio (the largest possible) by the incident particle charge-to-mass ratio. 
There are two essentially distinct situations. For an incident
electron, $D \simeq l_{\mathrm{c}} = n^{-1/3}$, as is natural, because
one electron is likely to lose a large fraction of its momentum in the
collision with another electron. In contrast, an incident 
atomic nucleus 
has a much smaller charge-to-mass ratio,  
so $D \gg l_{\mathrm{c}}$.
Indeed, the nucleus transfers a small fraction of its momentum 
in the interaction with one electron 
and has to interact with many electrons to lose its initial momentum.
%
Therefore, its motion is smooth and can be described by the differential
equation~(\ref{x-v}).
This description constitutes the {\em continuous slowing down
  approximation} (CSDA). 

To obtain an estimate of $D$ from (\ref{D/l}), we need to know
$l_{\mathrm{c}}$, that is to say, the interelectronic distance.
In condensed media, this distance is given by the atomic dimensions.
Even assuming that $D$ can be 
several thousand times larger than $l_{\mathrm{c}}$, 
the penetration depth of a fast atomic particle 
in a condensed medium should be macroscopically negligible.
This conclusion invalidates the hypothesis of a first-kind
asymptotics for large $v_0$ (even with the allowance for a logarithmic correction).

Let us instead assume asymptotic similarity, for $v_0 \gg v_{\mathrm{c}}$; namely, 
\begin{equation}
\tilde{D}(Q, v_0,  \rho_e, \rho_m) \sim
\frac{\rho_e^{2/3}}{Q^{2/3}\,
  \rho_m}\,\left(\frac{v_0\,\rho_m^{1/2}}{Q^{1/3}\rho_e^{2/3}}\right)^{\b}
\label{D-d}
\end{equation}
[which is just (\ref{D-powerlaw}) 
with all the exponents expressed in terms of $\b$ by solving 
Eqs.~(\ref{B1},\ref{B2},\ref{B3})].
We can place restrictions on $\b$ by reasoning
on how $\tilde{D}$ should depend on $Q, v_0,  \rho_e, \rho_m$. First,
it is necessary that $\b \geq 0$, if ${D}$ is not to shrink as $v_0$ grows.
Furthermore, $\b > 1$, if we assume that ${D}$ is to
shrink when $\rho_e$ grows and so does the interaction. The behavior of
${D}$ with respect to $\rho_m$ depends on whether $\b$ is smaller or
larger than two, but this is a moot point. 
A definite value of $\b$ is obtained by assuming that 
$\rho_e$
corresponds to a given type of constituent charges, which we naturally
take as electrons:
given that the electric force exerted by every electron
on the charge $Q$ is proportional to $eQ$ and $\rho_e= e\,n$, 
$\tilde{D}$ must be a symmetric function of $Q$ and $\rho_e$, which implies 
that $-2/3-\b/3=2/3-2\b/3$. Therefore, $\b=4$. 
This is the right value for a fast moving particle, 
as shown in detail in Sect.~\ref{model}. 

To find the slow-motion asymptotics, we can try keeping
the power law form (\ref{D-d}) and looking for a
first-kind asymptotics in some variable other than $v_0$. The condition 
$\b \geq 0$ forbids that $Q$ disappears from $\tilde{D}$ but 
we can try $\rho_m$ or $\rho_e$. The former disappears if $\b=2$,
and the latter if $\b=1$, making $\tilde{D}$ quadratic or linear in $v_0$,
respectively. These two cases correspond to stopping forces constant
or linear in velocity, as shown in Sect.~\ref{empiric}. A slow-motion
linear force, like in fluids, seems more plausible. 
Substituting $\b=1$ in (\ref{D-d}), we have
$\tilde{D} \sim v_0/(Q\,\rho_m^{1/2}),$
which implies
\begin{equation}
F \sim Q \rho_m^{1/2} v.
\label{FQ-lin}
\end{equation}
In Sect.~\ref{low-v}, we explain how such force arises.

To unveil the physical meaning of
$v_{\mathrm{c}}$, let us note that it is, according to Eq.~(\ref{vc}), the quotient 
of the length $l_{\mathrm{c}}=(Q/\rho_e)^{1/3}$ by the time $\rho_m^{1/2}/
\rho_e$. 
Instead of this time, we can consider its inverse, 
$\rho_e/\rho_m^{1/2}$, as a characteristic frequency
of a uniform charge distribution. 
This frequency corresponds to the
\emph{plasma frequency} $\omega_{\mathrm{p}}$ of the distribution of free electrons in a conducting medium,\cite{Feynman_2,Kittel}
namely,
\begin{equation}
\omega_{\mathrm{p}}^2=\frac{4\pi e^2n}{m},
\label{op}
\end{equation}
as seen by putting $\rho_e= e\,n$ and $\rho_m= m\,n$.  
The plasma frequency characterizes collective oscillations of the electrons 
about their rest state, which
can indeed absorb energy from a traversing particle.\cite{Feynman_2,Kittel} 
%
The free electron density $n$ of conducting media (on Earth) varies in a wide range, from $10^9$/m$^3$ (in the ionosphere) to $10^{29}$/m$^3$ (in metals).
Although we are not only concerned with conducting media, 
there is no definite distinction between conductors and 
dielectrics in their response to the rapidly varying electric field produced by a fast particle, because all electrons respond as free electrons to very
high frequency fields. 
To analyze the mechanism of energy loss, we need a detailed electrodynamic model. 

\section{Electrodynamic model of the resistance force}
\label{model}

The theory of the stopping of fast charged particles in matter is well established.
In the macroscopic theory, the energy of the particle is lost in polarizing the medium,
whereas in microscopic models the particle interacts with a distribution of mobile electrons. 
We rely on two classic textbooks of electrodynamics: Landau and Lifshitz's,\cite{L-L}
which concisely explains the macroscopic theory, and Jackson's,\cite{Jack} 
which focuses on the microscopic processes. 
For a recent and comprehensive reference on the subject, see Sigmund.\cite{Sigmund}

Before looking into the details, let us confirm that $\b = 4$ in the asymptotic law (\ref{D-d}).
Let us consider that the resistance force is just a self force, namely, a force exerted on the particle by its own electric field.\cite{L-L} 
Because the electric field of the particle is proportional to $Q$, 
$F$ must be proportional to $Q^2$. From this condition and Eq.~(\ref{D-int}),
we deduce that $\tilde{D}$ is inversely proportional to $Q^2$ and, 
according to (\ref{D-d}), 
\begin{equation}
\tilde{D} \sim
\frac{\rho_m\,v_0^4}{Q^{2}\rho_e^{2}}\,.
\label{D-2}
\end{equation}
Correspondingly,
\begin{equation}
F \sim \frac{Q^{2}\rho_e^{2}}{\rho_m v^2} = 
\frac{z^{2}e^{4}n}{m v^2}\,.
\label{F_nolog}
\end{equation}
These asymptotic laws are valid for fast particles, 
namely, for $v \gg v_{\mathrm{c}} =
l_{\mathrm{c}}\,\omega_{\mathrm{p}},$ 
which is related to Landau and Lifshitz's condition for their macroscopic treatment, namely,
$v \gg a\,\o_0$, 
where $\o_0$ is ``some mean frequency
corresponding to the motion of the majority of the electrons in the atom''
and $a$ is the interatomic distance.

\subsection{Calculation of the resistance force and the penetration depth}
\label{LL-J_calc}

Landau and Lifshitz\cite{L-L} express the self force as
an integral in Fourier space. For a point-like particle in the vacuum, this integral diverges for large wave numbers, but the result must vanish by symmetry (isotropy). This holds for a particle moving in a medium with constant electric permittivity, but, when the permittivity depends on the frequency (as it must), the divergent integral can be regularized to yield:\cite{L-L}
\begin{equation}
F(v) = \frac{4\pi z^{2}e^{4}n}{m\,v^2}\,\ln\frac{k_0\,v}{\bar{\o}}\,,
\label{F-LL}
\end{equation}
where $k_0$ 
is a transversal wave-number cutoff ($k_0 \ll a^{-1}$ 
for the macroscopic treatment to be valid),
and $\bar{\o}$ is a mean electronic frequency, 
defined in terms of the permittivity.
Eq.~(\ref{F-LL}) agrees with the law (\ref{F_nolog}),
up to the soft $v$-dependence in the logarithmic factor.
$F$ is actually due to the imaginary part of the 
permittivity, which represents the absorption 
of electromagnetic energy.\cite{Feynman_2,Kittel,L-L,Jack}  
This process is 
different in conductors and in dielectrics at low frequencies but
we can assume, for very fast particles, that 
$\o_{\mathrm{p}}\simeq \bar{\o}$ and that they both are
close to the basic atomic frequency, namely,
the Bohr frequency $me^4/\hbar^3$.


Jackson\cite{Jack} considers the stopping force as the result of the collisions of the incident particle with individual electrons. 
%
Assuming the CSDA to be valid, the stopping force 
produced by a uniform distribution of particles is equal to the energy 
transferred to them per unit length. 
The energy transferred to the particles in an element of volume, 
if the particles are independent, is given by 
the energy transfer $T$ to one particle times the number of particles,
$n\,dS\, dx$. 
Therefore, 
\begin{equation}
F = \frac{dE}{dx} = n\int T\,dS.
\label{F-T}
\end{equation}
We can identify this formula with Eq.~(\ref{F-Newt}).
Indeed, if $\D \bm{p}$ denotes the momentum transfer to one particle, 
$$T = \frac{Mv^2}{2} - \frac{(M\bm{v}- \D\bm{p})^2}{2M} \approx 
\bm{v}\cdot\D \bm{p} = m\,v\,\D v_{\parallel}\,.$$ 
It is to be remarked that collisions with free electrons are elastic and 
$T$ fully transforms into kinetic energy, so that
$T= (\D p)^2/(2m) = m[(\D v_{\parallel})^2 +(\D v_{\perp})^2]/2$, unlike in Sect.~\ref{drag}.
 
The collision is best studied in the rest frame of the incident particle.\cite{Jack}
The energy transfer for Coulomb scattering is
\begin{equation}
T(b) = \frac{2z^2e^4}{m v^2} \,\frac{1}{b^2 + b_0^2}\,,
\label{T_C}
\end{equation}
where $b$ is the {\em impact parameter} and $b_0 = z e^2/(mv^2)$.\cite{b0} 
Because $T(b) \neq c(b)\, mv^2$, the total stopping force, 
as given by Eq.~(\ref{F-T}), is not proportional to $v^2$, 
unlike in Newton's theory, Eq.~(\ref{F-Newt_1}). 
Writing $dS = 2\pi b\, db$ and
carrying out the integration over $b$, one finds that
\begin{equation}
F = 
n\int_{b_{\mathrm{min}}}^{b_{\mathrm{max}}} T(b)\,2\pi b\, db = 
2\pi n \frac{z^2e^4}{m v^2} \ln \frac{b_{\mathrm{max}}^2 +
  b_0^2}{b_{\mathrm{min}}^2 + b_0^2}\,.
\label{F-J_bb}
\end{equation}
In this context, the Coulomb interaction is of long range because $F$ diverges
if $b_{\mathrm{max}} \ra \infty$.

The value of $b_{\mathrm{max}}$ is determined by 
Bohr's adiabaticity condition:\cite{Sigmund,Jack} 
when the time of passage of the incident particle by one electron, of the order of $b/v$, is long compared to the characteristic period of its motion, 
the response of the medium is adiabatic, that is to say, reversible, and no energy is transferred (however, see Sect.~\ref{low-v}).
In the opposite case, the electron can be considered free and at rest. 
Therefore, we can take
\begin{equation}
b_{\mathrm{max}} \simeq v/\bar{\o}.
\label{bmax}
\end{equation}
If we further make $\bar{\o} \simeq me^4/\hbar^3$ (the Bohr frequency), 
the ratio of $b_{\mathrm{max}}$ to the Bohr radius, $a_0=\hbar^2/(me^2)$,
is approximately the ratio of $v$ to the Bohr velocity $e^2/\hbar$.
This ratio is bounded above by $\hbar c/e^2 = \a^{-1}$, 
the inverse of the universal constant $\a=e^2/(\hbar c) = 1/137$ 
(the \emph{fine-structure constant}). 
Therefore, $b_{\mathrm{max}} < a_0/\a = 137a_0$ 
(neglecting relativistic effects, of course). However, in a condensed medium, an atomic monolayer of linear dimension $100a_0$ 
contains of the order of $10^4$ atoms and more mobile electrons.

Let us now turn to $b_{\mathrm{min}}$ in Eq.~(\ref{F-J_bb}). 
For \emph{soft collisions}, such that the momentum transfer to every electron is small, 
that is to say, 
for $b_{\mathrm{min}} = 1/k_0 \gg b_0$, 
$$
F \approx 
4\pi n\frac{z^2e^4}{m v^2} \ln \frac{b_{\mathrm{max}}}{{b_{\mathrm{min}}}}
=
\frac{4\pi z^{2}e^{4}n}{m\,v^2}\,\ln\frac{k_0\,v}{\bar{\o}}\,,
$$
which coincides with Eq.~(\ref{F-LL}). If we keep $b_{\mathrm{max}} =
v/\bar{\o}$ but integrate down to $b_{\mathrm{min}} = 0$, including 
\emph{hard collisions}, the argument of the logarithm is larger, namely,
\begin{equation}
F \approx 
4\pi n\frac{z^2e^4}{m v^2} \ln \frac{b_{\mathrm{max}}}{{b_0}} =
\frac{4\pi z^{2}e^{4}n}{m\,v^2}\,\ln\frac{mv^3}{z e^2 \bar{\o}}\,.
\label{F-J_all}
\end{equation}

Hence, we obtain the penetration depth: 
\begin{equation}
\tilde{D} = \int_0^{v_0} \frac{v\,dv}{F(v)}
\approx  \frac{m}{4\pi z^{2}e^{4}n}
\int_0^{v_0} 
\frac{v^3\,dv}{\ln[mv^3/(ze^2\bar{\o})]}\,.
\label{li}
\end{equation}
The integration over $v$ cannot be carried out analytically, unless one
ignores the logarithmic factor, whence the integration is trivial and agrees
with (\ref{D-2}).  
The logarithmic factor actually makes the integral divergent,
because of the pole at 
$v = \left({z e^2 \bar{\o}}/{m}\right)^{1/3}.$
After identifying $\bar{\o}$ with $\o_{\mathrm{p}}$,
this velocity is essentially the $v_{\mathrm{c}}$ in Eq.~(\ref{vc}), 
as shown by making $Q=ze,$ $\rho_e= e\,n$,
$\rho_m= m\,n$, and using Eq.~(\ref{op}).
So let us also identify $v_{\mathrm{c}}$ with the value of $v$ at the pole.
Of course, Eq.~(\ref{F-J_all}) is only valid for $v \gg v_{\mathrm{c}}$, 
which means that the pole is irrelevant.
To calculate the integral, let us first make the change of variable 
$r = v^4/v_{\mathrm{c}}^4$. Then,
$$
\int_0^{v_0} 
\frac{v^3\,dv}{\ln(v^3/v_{\mathrm{c}}^3)}
= \frac{v_{\mathrm{c}}^4}{3}  
\int_0^{r_0} \frac{dr}{\ln r}\,,
$$
where $r_0 = v_0^4/v_{\mathrm{c}}^4 \gg 1.$ 
The integral over $r$ is 
the {\em logarithmic integral}.
Its asymptotic expansion for $r_0 \gg 1$ can be obtained by 
repeated integration by parts,
but it is only justified to take 
the first term, because (\ref{F-J_all}) is already an asymptotic formula.
Therefore, we write:
\begin{equation}
\tilde{D} 
\approx
\frac{m}{4\pi z^{2}e^{4}n} \,
\frac{v_{\mathrm{c}}^4}{3}  \,
\frac{r_0}{\ln r_0} = 
\frac{m v_0^4}{48\pi z^{2}e^{4}n\ln(v_0/v_{\mathrm{c}})} 
\,.
\label{D-I}
\end{equation}
This is the best result achievable, but it is sufficiently accurate. 
We can compare Eq.~(\ref{D-I}) with the general form
in Eq.~(\ref{D-f}) 
and deduce that $f(x) \sim {x^4}/{\ln x}.$

So far, we have neglected quantum effects, which appear when the distance of closest approach is smaller than the de Broglie wave-length of the electron.
If $b_0 < h/(mv)$, we should replace
$b_{\mathrm{max}}/b_0$ by $b_{\mathrm{max}}/[h/(mv)] = mv^2/(h \bar{\o})$ in
Eq.~(\ref{F-J_all}).
The exact value of the argument of the logarithm can be obtained with a semiclassical
treatment, including momentum transfers up to $\hbar
q_{\mathrm{max}} = 2mv$, 
which yields:\cite{L-L,Jack}
\begin{equation}
F(v) = \frac{4\pi z^{2}e^{4}n}{m\,v^2}\,\ln\frac{2mv^2}{\hbar\bar{\o}}\,.
\label{F-LL_1}
\end{equation}
%
%
If we define the new characteristic velocity $v_{\mathrm{q}}= [\hbar\bar{\o}/(2m)]^{1/2}$,
then the argument of the logarithm is 
$(v/v_{\mathrm{c}})^{3}$ in the classical case and $(v/v_{\mathrm{q}})^{2}$ 
in the quantum-mechanical case.
The transition from the first to the second takes place when 
$(v/v_{\mathrm{c}})^{3} \gtrsim (v/v_{\mathrm{q}})^{2}$, namely,
when $v \gtrsim ze^2/\hbar$.

The calculation of the penetration range corresponding to Eq.~(\ref{F-LL_1}) 
yields:
\begin{equation}
\tilde{D} = 
\frac{m v_0^4}{32\pi z^{2}e^{4}n\ln(v_0/v_{\mathrm{q}})} 
\,.
\label{D-I_1}
\end{equation}
The maximal penetration, for $v_0$ close to $c$ but before relativistic effects take hold, is
{\setlength\arraycolsep{2pt}
\begin{eqnarray*}
D_\mathrm{max} &=&
\frac{M m c^4}{16\pi z^{2}e^{4}n\ln[2mc^2/(\hbar\bar{\o})]} \\
&=& \frac{M}{16\pi z^2\,m \,n \, \a^4 a_0^2\,\ln [2mc^2/(\hbar\bar{\o})]}\,.
\end{eqnarray*}}%
Using the Bohr frequency for $\bar{\o}$, 
$\ln [2mc^2/(\hbar\bar{\o})] = \ln (2\a^{-2}) = 10$,
which is the maximal value of the logarithmic factor.
With this value, we obtain
\begin{equation}
\frac{D_\mathrm{max}}{a_0} 
= 7\cdot10^5\,\frac{M}{z^2\,m\,(na_0^3)}
\,. 
\label{Dmax}
\end{equation}
In a dense medium, $na_0^3 \simeq 1$. Then, 
for a proton, with $z=1$ and $M/m = 1800,$
$D_\mathrm{max} \simeq 1.3\cdot10^9\,a_0 \simeq 0.1$~m. 
For a relativistic particle, ignoring the small change of the logarithmic factor, $F=40\pi z^{2}e^{4}n/(mc^2)$ and penetration depth is proportional to kinetic energy.

\subsection{Low-velocity particle}
\label{low-v}

A particle with $v<v_{\mathrm{c}}$, that is to say, with 
$b_{\mathrm{max}}< b_0$, 
adiabatically decouples from the electrons in the medium. 
However, there is a residual ``viscosity'' force of the form (\ref{FQ-lin}),
due to the interaction with the least bound electrons.  
The precise form of $F(v)$, according to Gryzi\'nsky,\cite{Gry} is given by 
the force produced by the electrons with velocity $v_{\mathrm{e}}$ 
on a particle with velocity $v$
integrated over the electron velocity distribution $n(v_{\mathrm{e}})\,dv_{\mathrm{e}}$.
In turn, the force produced by the electrons with velocity $v_{\mathrm{e}}$
is given by an integral over the relative velocity with respect to the particle. 
Gryzi\'nsky calculated the latter integral 
and obtained a complex but explicit formula.\cite{Gry} However, this formula is not suitable for integrating over $v_{\mathrm{e}}$.
%
%
%
One can proceed by splitting the integral over $v_{\mathrm{e}}$ into the 
regions $v_{\mathrm{e}} < v$ and $v_{\mathrm{e}} > v$ and using 
approximations of the integrand for, respectively, 
$v_{\mathrm{e}} \ll v$ and $v_{\mathrm{e}} \gg v$ 
(obtained from Gryzi\'nsky's formula or directly from the integral over relative velocities).
If $v$ is beyond the range of $v_{\mathrm{e}}$ [the support of $n(v_{\mathrm{e}})$] or, at any rate, is much higher than $\bar{v}_{\mathrm{e}}$
(the rms $v_{\mathrm{e}}$), then
the integral in the region $v_{\mathrm{e}} > v$ is negligible while the 
integral in the region $v_{\mathrm{e}} < v$ recovers Eq.~(\ref{F-J_all}).

%
%

Of course, we are now interested in the opposite case, namely, 
$v \ll \bar{v}_{\mathrm{e}}\,.$ 
It yields a force proportional to $v$, in accord with (\ref{FQ-lin}),
but it contains the factor $f(\bar{v}_{\mathrm{e}}/v_{\mathrm{c}})$, with
$f(x) \propto x^{-3}\ln x$.
Notice that the exponent $-3$ implies that $F$ is
actually proportional to $Q^2\rho_e^2$, in spite of appearing to be just proportional to $Q$ in (\ref{FQ-lin}). 
The value of $\bar{v}_{\mathrm{e}}$ is given by the properties of the medium, 
in terms of the type of electronic distribution. 
For the Maxwell-Boltzmann distribution of a classical plasma, 
$\bar{v}_{\mathrm{e}}$ is given by the temperature. 
For the Fermi-Dirac distribution
of a degenerate electron gas, which is more relevant in our context, 
$\bar{v}_{\mathrm{e}}$ can be replaced by the Fermi velocity,
given by the total electron number density $n$. Then $F$ follows
the Fermi-Teller formula,\cite{Gry} which, curiously, is essentially independent of $n$. Other formulas for $F$, namely, 
the Firsov and the Lindhard-Scharff formulas,\cite{ion-solid,Sigmund} 
derived in terms of low-$v$ particle-atom collisions, 
yield proportionality to $n$, 
but they qualitatively agree with the Fermi-Teller formula if  
$n \simeq a_0^{-3}$.

The argument of the logarithm for high $v$, namely, $v/v_{\mathrm{c}}$, is replaced by $\bar{v}_{\mathrm{e}}/v_{\mathrm{c}}$ for low $v$, which 
suggests that $\bar{v}_{\mathrm{e}}/\o_{\mathrm{p}}$ is the effective range 
in the low-$v$ limit.
In fact, this length reproduces the screening length of a static electric charge, namely, either the Debye screening length of a thermal plasma or the Thomas-Fermi screening length of a Fermi gas.\cite{Kittel}
If the electron velocity and frequency take the Bohr values, 
then the screening length is the Bohr radius. 
Therefore, a slow atomic nucleus in a cold medium 
actually becomes a positive ion and eventually a neutral atom. 
This occurs in a range of $v$ given by the orbital speeds 
of the electrons in the ion (Bohr's screening criterion).\cite{Sigmund,ion-solid}

A slow-moving particle transfers energy not only to the electrons but also 
to the atomic nuclei. At high velocity, $F$ is inversely proportional to 
$mv^2$, where $m$ is the mass of the interacting particles of the medium, 
so $F$ is certainly due to the electrons. But  
$F$ is maximal at $v\simeq v_{\mathrm{c}} = 
\left({z e^2 \bar{\o}}/{m}\right)^{1/3},$
and becomes proportional to $m^{1/2}v$ for lower $v$
($F$ always depends on $m^{1/2}v$, in accord with Sect.~\ref{fund}).
For decreasing $v<v_{\mathrm{c}}$, the much smaller force due to the nuclei 
eventually takes over, 
because it is still inversely proportional to $mv^2$, 
where $m$ is now the nucleus mass. This occurs at $v \simeq 500$ km/s.
However, in that range of $v$, 
the charge is screened,  
turning the long-range Coulomb interaction into a short-range interaction.

\section{Nano-projectiles}
\label{nano}

As seen above, the velocity of the particle determines the type and range of its interaction with the medium. 
In particular, as a slow atomic nucleus
becomes a neutral atom, its initial charge $Q$
vanishes and its initial negligible size grows to the atomic size, which 
is not negligible. The variable
$\rho_e$ loses significance and, at the same time,
the total mass density of the medium becomes the relevant density, 
because most of the energy and momentum of the projectile are transferred to 
the nuclei rather than to the electrons.
Moreover, the reduced interaction range 
makes the interaction similar to a macroscopic contact repulsive interaction. 
Therefore, the parameters of a sufficiently 
slow nucleus are the same macroscopic parameters employed in Sect.~\ref{fast}.
There must be a transition, ruled by 
the screening process, from the linear force law for 
$v\lesssim v_{\mathrm{c}}$ to a quadratic force for small $v$. 

In fact, 
the force produced by atomic hard-sphere collisions is quadratic in $v$, 
according to the CSDA and Eq.~(\ref{F-T}), because 
$T=(\D{p})^2/(2m)$ and $\D{p} \propto mv$ for a hard-sphere collision, although it depends on $b$ ($m$ is the reduced mass). However, 
the real atomic repulsion has short but non-vanishing range, and
the dependence of $\D{p}$ on $v$ deviates from linearity. 
A better model of atomic repulsion is the power-law potential
$V(r) = V_0 \left(a/r\right)^s$, already employed in Sect.~\ref{hyperv}. 
In general, $\D p = 2p \cos \ff$, where $\ff$ is the angle between the direction of the incoming particle and the direction of the closest approach point.\cite{L-L_M} This angle is given by an integral over the distance $r$,
which can be expressed for the power-law potential in terms of the nondimensional 
variable $u=(a/b)^s\,V_0/(mv^2)$, so that $T = 2mv^2 \cos^2\!\ff(u)$. 
The change of integration variable from $b$ to $u$ in Eq.~(\ref{F-T}) gives the form of $F$, namely, 
$$F \propto n\,(mv^2)^{1-2/s}\,V_0^{2/s}a^2.$$
Notice that $F$ is linear in the number density $n$ but not 
in $\rho_m$; but, as $s$ grows and the potential gets harder,
$F$ tends to be linear in $\rho_m$ and actually becomes the quadratic Newton force.

Let us consider a realistic interatomic potential, namely,  
the ``universal'' ZBL potential, which provides a good description  
in a broad range of interaction energies.\cite{ion-solid}
Although it can only be handled numerically, 
an accurate analytical fit of the resulting $F(v)$ is available.
For small $v$, $F \sim (mv^2)^{1-0.21},$ which agrees with 
the result for the power-law potential with $s=9.4$. 
$F(v)$ has a maximum for a value of $v$ that depends on the 
atom types and is in the range of hundreds of km/s.
For large $v$, $F \sim v^{-2} \ln v$, due to the nuclear repulsion.
However, then there are, in addition to nuclear interactions, 
the nucleus-electron interactions that give rise 
to electronic viscosity. 
The dynamics in the velocity range between the maximum of the short-range atomic force and the absolute maximum of the total force is complex.

A projectile that consists of a cluster of atoms (or even a single atom)
is called a nano-projectile. 
If we assume that the cluster is spherical and its collisions with the 
atoms of the medium are elastic, we can relate the collision potential 
to the atom-atom collision potential. In particular, we can use 
$V(r) = V_0\,a^{s}\left(r-R\right)^{-s}$ for $r>R$ and $\infty$ for $r\leq R$, where $R$ is the cluster radius. 
As is natural, the collisions are harder for larger $R$, in a relative sense. 
To obtain a quantitative result, 
we can approximate the preceding $V(r)$ by a power law, by means of a linear expansion of $\ln V$ at $r=R+a$ in terms of $\ln r$, which yields 
$$V(r)/V_0 \propto \left(a/r\right)^{s(1+R/a)}$$
(the error is less than 10\% for $0.2 < V/V_0 < 5$). 
Therefore,  for $R \gg a$,
$$F \sim v^{2-\b}, \hbox{ with } \b \sim 1/R.$$
This model can be generalized to non-spherical clusters, provided that 
the cluster size is much larger than the atomic force range, and can also be generalized to collisions with molecules instead of atoms.

However, the assumption of elastic collisions is questionable: the collisions 
can induce relative motions of the atoms in the cluster. For small clusters, 
this process is similar to the excitation of molecular vibrations by the collisions of molecules in a hot gas.\cite{Zel,Ander} A vibration of frequency $\o$ is not excited while $mv^2/2 < \hbar\o \ll 1$ eV. 
Even for $mv^2 \gg \hbar\o$, the probability of excitation is small, 
as long as the collision is quasi-adiabatic, namely, as long as $v \ll a\,\o$ (which is a few percent of the Bohr velocity). At larger velocities,  
the atomic configuration of the cluster can change and, when
the collision energy reaches the cohesive energy per cluster 
atom, one of them can be knocked off.

For clusters with many atoms and many possible vibration modes, 
the collision is best described as an impact on a quasi-macroscopic cluster, although the velocity and ``density'' of the projectile are insufficient for
penetration and the result is just inelastic scattering. 
As the force between cluster and atom of the medium, 
we can employ 
$$\bm{f} = -V'(r)\,\bm{r}/r - g(r)\bm{v},$$ 
with a friction term to account for the inelasticity [$g(r)>0$].
To find the effect of a collision ruled by this force, 
we need to calculate the longitudinal transfer of momentum,
$\D p_{\parallel} = m\,\D v_{\parallel}$. Unfortunately, we cannot do this analytically, except in very simplified cases, e.g., 
in a head-on collision and with restrictions.\cite{grains} 
At any rate, this simplified calculation shows that $\D v_{\parallel}$ 
is not proportional to the initial relative velocity $v$ and
Newton's theory fails. In general, the friction term 
is negligible for low $v$ and the collision is elastic; whereas, during a collision at high $v$, the friction dominates until 
the relative velocity becomes low enough for the conservative part of 
$\bm{f}$ to come into action, 
but by then most of the kinetic energy has been dissipated.
For even higher $v$, the collision becomes, in practice, 
totally inelastic, 
and $\D v_{\parallel} \approx v$, in accord with Newton's theory (see Sect.~\ref{physics}). However, energetic collisions 
cause atom displacements (plasticity) and sputtering (ablation). 

Although we have considered inelasticity as a sink of 
kinetic energy, it can also be a source, because a collision can 
transform cluster internal energy into kinetic energy. 
For example, molecular vibrations in a hot gas are in thermodynamic equilibrium with the Maxwell-Boltzmann velocity distribution.\cite{Zel,Ander} 
In our case, the cluster sees an essentially unidirectional velocity distribution. 
Nevertheless, in a stationary situation, the transitions between vibrational states are ruled by a \emph{master equation}, and therefore the distribution of these states evolves towards a 
stationary distribution.\cite{Reif} In this dynamical equilibrium, the 
internal energy of the cluster concentrates on its frontal part.
The kinetic energy can grow in an individual collision ruled by the  dissipative force $\bm{f}$
because $\bm{f}$ must be complemented with a stochastic or fluctuating component, in accord with the fluctuation-dissipation theorem. Therefore, the result of a collision is not determined by the initial conditions, 
and $\D v_{\parallel}$ is a random variable, with a stationary probability distribution in the dynamical equilibrium.
However, 
if $\D v_{\parallel} \propto v$ for the corresponding elastic collision,
also $\langle\D v_{\parallel}\rangle \propto v$.


At any rate, the hypothesis of binary encounters between the cluster and the 
atoms of the medium only applies to small clusters, in the following sense.
We recall from Sect.~\ref{physics} that the transition from the continuum to the molecular-kinetic description of penetration in a fluid is determined by the ratio of the body length $l$ to the molecular mean free path $\l$.
Given that a solid medium responds as a fluid
in penetration at high $v$, the same criterion applies.
Notice that the mean free path is independent of the temperature and 
only depends on the number density and interaction cross-section
of the molecules of the medium.\cite{Reif} 
When $a \ll l \ll \l$, the binary-encounter model and the law $\b \sim l^{-1}$
hold. However, when $l \gg \l$, the cluster affects simultaneously many 
atoms, justifying an approach based on 
the hypothesis of {\em local} thermal equilibrium.\cite{Reif}
Then, the nano-projectile generates
a mesoscopic version of the macroscopic bow wave studied
in Sect.~\ref{hyperv}. In solids, it is normal to describe 
mesoscopic penetration in terms of a
{\em thermal spike},\cite{ion-solid}
but it can also be described in terms of a shock wave.\cite{Insepov}

Macroscopic projectiles in rarefied gases can also be ``small''
in comparison with the corresponding molecular mean free path; 
e.g., meteorites or spacecraft at high altitudes.
In this context, the preceding issues, namely, type of drag, 
conversion of kinetic to internal energy, ablation, etc, have been much 
studied.\cite{Bronshten,Trib}
In steady hypersonic motion, $F$ is quadratic in $v$ and the heat transfer to the body, as a fraction of the power $Fv$, is proportional to $\rho_{\mathrm{m}}v^3$. However, at altitudes 
such that the molecular mean free path is macroscopically large,
$\rho_{\mathrm{m}}$ is so small that the heating is not significant. 
Nevertheless, cold sputtering can be significant in long duration spacecraft missions.\cite{Trib} Of course, 
thermal ablation is important at lower altitudes, for meteorites or for spacecraft at reentry (its major effect on meteorites is mentioned in 
Sect.~\ref{hyperv}).
In the case of atom clusters, 
the energy transfer is also of thermal nature, provided that 
local thermal equilibrium holds inside the cluster 
(a condition less strict than that it hold outside). 
Needless to say, a cluster penetrating in a dense medium can reach
high temperatures, although the effects of stress are surely more important (Sect.~\ref{hyperv}).

\subsection{Critique of reports of molecular dynamics simulations}

Most studies of atom cluster penetration in solids rely on 
molecular dynamics simulations. Employing this method, 
some authors\cite{Carr} conclude that $D$ is quadratic in $v_0$ and
others\cite{Prat} that it is linear, corresponding to a constant or 
linear force, respectively (Sect.~\ref{empiric}).  
However, their data fits are questionable
(e.g., Fig.~3 of Ref.~\onlinecite{Carr} and  Fig.~3 of Ref.~\onlinecite{Prat}). 
Moreover, the penetration depths only reach a few nm.
The simulations by Anders and Urbassek\cite{Anders_cw,Anders_cohesive}
of the stopping of heavy clusters in soft targets (Au clusters in solid Ar) 
reach depths of tens of nm.
These simulations span a range of energy per cluster atom between 
10 and 1000 eV, equivalent to a range of $mv^2/2$ between 2 and 200 eV. 
Measures of the initial force find that it 
is nearly linear in $v^2$. To be precise, the power-law 
exponent of $v^2$ grows with cluster size, being, e.g., $0.87$
for Au$_{43}$ and $1.01$
for Au$_{402}$.\cite{Anders_cw} 
For given $v$, the force depends on the number of atoms in the cluster
as $N^{2/3}$; 
that is to say, it is proportional to the cross-sectional area (although the exponent of $N$ grows with $v$ and tends to 1 for large $v$).

The simulations of self-bombardment (with cluster and target atoms of the same type) by Anders {\em et al}\cite{Anders_NJP} obtain an $F$ versus $v^2$ exponent $\a(N)$ that starts at 0.6 for $N=1$ and tends to 1 for large $N$.
Although self-bombardment does not produce deep penetration, it is fine for studying the initial force, and these results for $\a$ agree with the 
preceding results in Ref.~\onlinecite{Anders_cw}.
Anders {\em et al}\cite{Anders_NJP} state that ``no theoretical argument exists regarding why the interaction of a cluster with a solid can be described by an energy-proportional stopping for large cluster sizes.'' Our model predicts precisely that 
the exponent $\a(N)$ is, for $N=1$, 
sensibly smaller than one ($\a \lesssim 0.8$) 
and tends to one for large $N$. Furthermore, the model (in a certain approximation) predicts that the difference, $1-\a$, decreases as $N^{-1/3}$.

Finally, let us comment on a theoretical analysis of cluster implantation 
made by Benguerba.\cite{Benguerba} This analysis 
has broad scope, in spite of being restricted to graphite targets.
Benguerba assumes that the dynamics is ruled by the displacement of target atoms caused by a ``generated wave,'' so that,
if $c_{\mathrm{t}}$ is the wave speed,
the medium mass $dm = \rho_{\mathrm{m}} A\,c_{\mathrm{t}} dt$
is displaced with the velocity $v$ of the projectile, and hence
$F = v\,dm/dt = A\rho_{\mathrm{m}}c_{\mathrm{t}}v$; that is to say, the force is linear. 
Although Benguerba does not specify the type of generated wave, it must be 
a shock wave, and his model indeed corresponds to the generation 
of a {\em weak} plane shock wave by an impulsive load and its 
propagation at a speed close to the speed of sound.\cite{Zel} 
However, a penetrating supersonic projectile generates a bow wave
with a shock front that propagates at the same velocity of the projectile. 
Therefore, $c_{\mathrm{t}}$ must be identified with $v$ in Benguerba's 
formula and $F$ is actually quadratic in $v$.

\section{Summary and Conclusions}

We have presented a unified formalism for projectile penetration that encompasses the full ranges of projectile sizes and velocities. 
As regards macroscopic projectiles,
Newton's inertial force of resistance, with its quadratic dependence 
on velocity, constitutes a universal law, 
because it only contains the basic mechanical variables, namely, the size and velocity of the projectile and the density of the medium. 
In correspondence with Newton's force, the universal formula for penetration depth is Gamow's formula. 
However, Gamow's formula {\em does not} follow from Newton's force. 
The formula that actually follows from Newton's force
contains a logarithm of the velocity and, therefore, 
introduces a critical velocity, 
showing that Newton's inertial force is only valid 
for fast projectiles.

The critical velocity must involve some variable that contains the time dimension and is associated to the mechanism of resistance at low velocity.
The resistance of incompressible fluids turns from viscous to inertial 
as the Reynolds number grows.
Compressible fluids or solids offer inertial resistance to supersonic 
motion, which involves bow waves. 
At velocities of some km/s, the integrity of macroscopic projectiles can be 
seriously compromised by the effects of high pressures, in condensed media, 
or by the effects of high temperatures, at somewhat larger velocities, 
in any media. 
The maximal velocity depends on several factors, such as the strength and shape of the projectile, but it is generally lower than 10 km/s. The lower end of the quadratic force range is very variable, especially, in fluids, which offer subsonic inertial resistance. 
The key factors for deep penetration are, first, a large ratio of projectile density to medium density and, second, 
a streamlined shape. The impact velocity is not important in the inertial regime, but the projectile must be strong enough to withstand 
the generated pressure and temperature.

The stopping of fast charged subatomic particles in matter is due to the long-range Coulomb force. The dimensional analysis of the most basic model 
unveils, besides a characteristic length, a built-in characteristic velocity, 
and hence a characteristic frequency, which can be related to 
the plasma frequency. Assuming asymptotic similarity 
for fast motion and considering the stopping force as an electric 
{\em self force}, we can determine its general form,
in particular, that it is inversely proportional to $v^2$, and hence 
that the penetration depth is proportional to $v_0^4$. These velocity dependences miss a logarithmic factor, which can be derived only with
detailed electrodynamic models. A macroscopic model that
includes the electric permittivity of the medium 
yields the logarithmic factor, with a mean frequency $\bar{\o}$ that replaces 
$\o_{\mathrm{p}}$.
An atomistic model, which considers
the transfer of momentum to electrons down to the smallest scales, 
only produces a slightly larger logarithmic factor
(generally smaller than 10 for non-relativistic particles). 

The characteristic velocity $v_{\mathrm{c}}=(ze^2\bar{\o}/m)^{1/3}$
is, in practice, of the order of magnitude of the Bohr velocity 
($e^2/\hbar = 2\hspace{1pt}200$ km/s). 
Projectiles with velocities in the range 10~--~2000 km/s are in 
the transition from the inertial fluid-like resistance described by 
the parameters $M, l, \rho_{\mathrm{m}}$, and caused by short-range interactions with the atoms of the medium
to the electric resistance described by the parameters $M, Q, \rho_e, \rho_m$, and caused by long-range interactions with the electrons of the medium.
In both cases, the resistance force arises
from the continuous transfer of momentum from the projectile to the medium,
and Eq.~(\ref{F-Newt}) is always valid.
The initial manifestation of the long-range electronic interaction, 
for velocities in the range 100~--~2000 km/s, is as 
an ``electronic viscosity'' force.
Naturally, when $v$ approaches 2000 km/s, the projectile can only consist of 
one bare atomic nucleus. 
Indeed, an atom cluster at that velocity rapidly fragments and
the resulting atoms are fully ionized.
The force on an atom cluster in the range 1~--~100 km/s is basically given by 
the inertial Newton's law, as long as the cluster is stable.
However, for nano-projectiles with diameter $l$ smaller than the mean free path of the particles of the medium, 
the force is not exactly quadratic but is
$F \sim v^{2-\b}$, with $\b$ small and vanishing as $l^{-1}$. 

From a fundamental standpoint, the resistance to
a projectile traversing a medium takes place because of the
irreversible transfer of momentum (or kinetic energy) to the medium. 
The main cause of it is a non adiabatic interaction.
For example, non-adiabaticity manifests itself as an increase of entropy in the bow wave generated by a supersonic projectile. 
The non-adiabatic interaction of a fast charged particle manifests itself in the irreversible transfer of energy to the high-frequency modes of
electronic motion, which results in a resistance force that is large but is such that it decreases with velocity, allowing for considerable penetration depths. 
For velocities below the Bohr velocity, the electronic interactions become adiabatic, but there can be non-adiabatic transfer of energy to low-frequency 
atomic motions. 
Nevertheless, there is also ``electronic viscosity.'' This type of viscosity as well as the ordinary viscosity in fluid motion are forms of 
irreversible transfer of momentum that belong to the general class of transport processes. 

A few final remarks of practical nature. Gamow's formula is, in fact, a sensible and quick rule for
estimating the penetration depth of a fast projectile of generic shape. 
For non-relativistic subatomic particles, formula (\ref{Dmax}) gives 
the maximum penetration depth, and
the penetration depth for a given velocity $v_0$ is obtained 
from it by multiplying it by $(v_0/c)^4$
(ignoring the small change of the logarithmic factor). The penetration depth of relativistic particles is proportional to their kinetic energy.




\begin{thebibliography}{99}
\bibitem{shock-dissip}
The specific energy transfer $vv_{\parallel}$ is converted to 
kinetic energy $(v_{\parallel}^2 +v_{\perp}^2)/2$ and internal energy.
The boundary conditions at the shock front (Ref.~\onlinecite{L-L_FM}) imply 
that the medium behind it moves perpendicular to it and that its maximum 
velocity is $v \sin{\a}$, where $\a$ is the angle of the shock front with the 
longitudinal axis, that is to say, $\tan{\a} = v_{\parallel}/v_{\perp}$.
Therefore, the minimum fraction of energy converted to internal energy  is $1 - (v_{\parallel}^2 +v_{\perp}^2)/(2vv_{\parallel})  = 1/2$, for any $\a$.
\bibitem{b0} 
In the collision of charges with the same sign, 
$b_0 = z e^2/(mv^2)$ is half the distance of closest approach in a head-on collision with velocity $v$.
\end{thebibliography}
\end{document}